\documentclass[a4paper]{article}                     
\usepackage{graphicx}
\usepackage{amsfonts}
\usepackage{amsmath}
\usepackage{amssymb}

\usepackage{epsfig}
\usepackage{psfrag}

\begin{document}

\title{Numerical modelling of unsaturated-saturated flow under centrifugation with no outflow}
%
\author{J.\ Ka\v{c}ur\\
          Faculty of Mathematics, Physics and Informatics,\\ 
          Comenius University Bratislava\\
          Slovakia \\
          \texttt{kacur@fmph.uniba.sk}
\and B.\ Malengier\\
            Department of Mathematical Analysis, Research Group \textit{NaM}$^2$,\\
            Ghent University,\\
            Galglaan 2, B-9000 Ghent, Belgium\\
            Tel.: +32 9 264 49 53\\
            Fax: +32 9 264 49 87\\
	    \texttt{bm@cage.UGent.be}
\and P.\ Ki\v{s}o\v{n}\\
          Faculty of Mathematics, Physics and Informatics,\\ 
          Comenius University Bratislava\\
          Slovakia \\
          \texttt{kisson@fmph.uniba.sk}}

\date{November 1, 2009}
%
\maketitle
\begin{abstract} 
A novel centrifuge set-up for the study of unsaturated flow characteristics in porous media is examined. In this set-up, simple boundary conditions can be used, but a free moving boundary between unsaturated-saturated flow arises.
A precise and numerically efficient approximation is presented for the mathematical model based on Richards' nonlinear and degenerate equation expressed in terms of effective saturation using the Van Genuchten-Mualem ansatz for the soil parameters in the unsaturated zone. Sensitivity of the measurable quantities (rotational moment, center of gravity and time period to achieve quasi steady state) on the soil parameters is investigated in several numerical experiments. They show that the set-up is suitable for the determination of the soil parameters via the solution of an inverse problem in an iterative way, excluding the saturated hydraulic conductivity. For this parameter, an existing simple centrifuge set-up is repeated and augmented with transient measurements.

\end{abstract}

\section{Introduction}
To predict the flow and solute transport in soils one needs the soil hydraulic
properties in terms of soil parameters. Once determined, these parameters can be used as input data in the governing mathematical model. For unsaturated flow, this model is given in terms of the saturation and the pressure head in Richards' equation (see below), which is a  nonlinear and degenerate parabolic equation. Furthermore, when part of the sample is saturated, free boundaries between the saturated zone and the partially saturated zone arise, as well as between the dry and the partially saturated zone. This is a major problem for many modeling approaches, leading to experimental set-ups that avoid the formation of these boundaries.

The soil retention and hydraulic permeability functions linking the saturation and pressure head for  unsaturated flow are expressed using the Van Genuchten-Mualem ansatz by means of soil parameters. Measuring these soil parameters is 
usually time consuming and tedious, especially for low conductive porous media. Several set-ups based on centrifugation have been proposed to obtain a large acceleration of the processes involved, see \cite{ConWri,NimAks,[NM],[SN],Den} and citations therein. These techniques have several disadvantages. An approach which aims for a steady-state flow regime inside the centrifuge, \cite{ConWri,NimAks}, requires expensive and/or complex apparatus, and obtains only a few water content versus conductivity measurements per run. Also transient set-ups based on keeping a top boundary at a fixed prescribed setting, \cite{Den}, are expensive. The quasi-steady centrifuge (QSC) method, \cite{CapNim}, is a much simpler technique (a slowly emptying reservoir at the top that is refilled when needed), but requires that the criterion for steadiness of flow through the sample is relaxed, leading to higher uncertainty in the obtained results. 

The alternatives for determining conductivity with a steady-state flow, couple a transient flow with parameter estimation techniques, see eg.\ \cite{Den,[SN]}. In this way, the conductivity and retention curve can be determined inversely over a large saturation domain. These methods require experiments of some state variables which relate to the conductivity. One-step or multi-step outflow methods are common in column experiments. Pressures or suctions are applied at the top or the bottom of the sample, while the outflow, and optionally pressure head within the sample, are measured. These measurements are then used to estimate the hydraulic parameters. This technique is transferred to the centrifuge device in \cite{[SN]}. However, measuring precise outflow rates is problematic in a centrifuge. Therefore, their method for determination of the soil parameters  is based on transient water content measurements (coming from electrodes installed in the specimen). This can be used in an equilibrium analysis as well as in a transient analysis. In this setting, the specimen is fully saturated at the beginning, and saturation is maintained at the outflow via a short lip at the end of the container. Good results are obtained, but there remain some disadvantages to this technique: there are few measurements close to saturation, leading to a high error in the prediction of the conductivity close to saturation, the sample needs to be disturbed to introduce the electrodes, and there is a very long waiting time in order to achieve the equilibrium when the equilibrium analysis approach is used.

The main goal of this manuscript is to develop a precise numerical method enabling to determine the soil parameters (via solution of inverse problem) in a very simple way requiring very cheap measurements.

In this article we consider any partially saturated sample which is sealed at the right boundary (from the center of centrifuge) and has no inflow at the left boundary. The amount of infiltrated water is fixed and would be important initial information, but in the fact, we can drop this requirement and consider the initial water content as an additional parameter requiring determination.
The only measurements required in our setting is the rotational momentum and the center of gravity of the sample in a set of equilibria corresponding to some predetermined rotational speeds. This can be replaced however by with transient measurements of these variables allowing to reduce the centrifuge operating time. These measurements are sufficient due to the fact that the saturation profile at the equilibria is not depending on the initial distribution of water in specimen, but only on its amount, which, when the right boundary of the sample is sealed, is identical in all equilibria. 

To use this procedure, we have to face serious difficulties in the numerical modeling. The main one is that if the right side of the sample reaches effective saturation, an interface between partially saturated zone and saturated zone appears. This boundary is very difficult to control numerically, causing problems with the mass balance conservation which is very important in this set-up.

To reach the equilibrium is an infinite asymptotic process, but after some time (eg. 1-3 days for low conductive material) the change of the rotational momentum and of the center of gravity can no longer be measured. At that moment, the rotational speed is increased, and the system moves towards a new corresponding equilibrium. Note that even when equilibrium was not reached and a small error is present in the measurements of the rotational momentum and the center of gravity, this will not influence the error at the higher equilibrium level. This error depends only on the running time of centrifugation at the actual rotational speed. The differences between applied rotational speeds are chosen in such a way that that the differences in outputs (rotational momentum and center of gravity) are technically well distinguishable. 

Next, the soil parameters and eventually the amount of originally infiltrated water, can be determined by minimizing a cost functional expressing the distance between the measured and the computed output, e.g., with the Levenberg-Marquard method. In this manu\-script we only show the sensitivity of the outputs on the parameters, which is a good indication of how an inverse method will perform. In fact, the correct determination of soil parameters substantially depends on the sensitivity and the errors in practical measurements.

We can also use the transient measurements of the outputs, for example the period needed to cross from one equilibrium to another. This would give us additional information useful in the determination of the soil parameters.

The advantage of the above approach is that the full range of saturation values are present in the setup, while preventing outflow means equilibrium can be obtained faster. However, due to the set-up, it is clear that the water flows from the unsaturated zone to the saturated zone, with no flow occurring in the saturated zone. Indeed, we notice that the rotational momentum and center of gravity in the equilibria approach are not sufficiently sensitive on the ``saturated hydraulic conductivity''. Therefore, we repeat in Section \ref{sat} also centrifugation with a fully saturated specimen which substantially increases sensitivity on this soil parameter, see \cite{[NM]}. In this setup the sample remains saturated and the water table head in the top reservoir is measured over time and related to the saturated hydraulic conductivity. It is in other words the same as a column experiment where the dropping head at the top is measured instead of the outflow, and the head at the bottom is fixed instead of the head at the top.

In our numerical method we reduce the mathematical model to a system of ODE using method of lines (MOL), which has already been successfully applied to Richards' equation in e.g.\ \cite{TocKel}. Our main contribution is in correctly handling the moving free boundary. The obtained system can be solved with ODE solvers for stiff systems and stiff DAE systems (system of ODE and algebraic equations). The numerical experiments demonstrate that this solution method is numerically very efficient (precise and economical). Note that reaching equilibrium for a high rotational speed can require a week or even a month of centrifugation for low conductive materials, whereas the computational time is only a few seconds. Therefore this method is a good candidate for solving the inverse problem, as this requires many computations of the forward problem. 
The numerical method can be successfully applied in other centrifugation settings (concerning control of the inflow, or control of the outflow) as, e.g., in \cite{[SN],Den}.

The research performed in this article was done for a preliminary study for the feasibility of a new centrifuge device for low conductive materials that should be cheap to operate.  The results presented here indicate that an approach where only rotational moment and gravity center is needed, is workable. This would remove the need to add experimental apparatus to the centrifuge measuring pressure heads, water contents or outflow rates during the operation. The authors intend to investigate a different set-up in a future article, in order to determine an optimal candidate, as well as collaborate to create a prototype centrifuge.
 
In Section \ref{mathmod} we present the mathematical model, giving specific attention to the movement of the free boundary. In Section \ref{nummeth} the numerical method based on the MOL approach is given, while in Section \ref{sat} the approach to determine the saturated hydraulic conductivity is explained. We finish in Section \ref{numexp} with several numerical experiments showing the sensitivity of the output parameters on the soil parameters.

\section{Mathematical model}\label{mathmod}
We consider a one dimensional model for a partially saturated sample in the form of a tube. The tube starts (top or left boundary) at the distance $r=r_{0}$ from the center of the centrifuge and ends at the distance $r=r_{0}+L$. The right boundary of the specimen is isolated. 
Flow in porous media under centrifugation is modeled by Darcy's equation in the saturated region and  by Richards' equation in the unsaturated region (see, e.g., \cite{[SN]},\cite{Den}). So
 \begin{equation}
\partial_r \left[K_s\left(\partial_r h- \frac{\omega^2}{g}r\right)\right]
=0 ,\label{eq1.1a}
\end{equation}
in the saturated region, and 
\begin{equation}
 \partial_t \theta = \partial_r \left[k(\theta)\left(\partial_r h- \frac{\omega^2}{g}r\right)\right],
\label{eq1.1b}
\end{equation}
in the unsaturated region. Here, $h$ is the piesometric head,  $\theta$  the saturation of the porous medium, $\omega $ the angular speed of rotation (in radians per second), $K_s$ the hydraulic conductivity in the saturated region, $g$ the gravitational constant 
and the function $k(\theta)$ describes the hydraulic conductivity in the unsaturated region. Denote by $u= \frac{\theta-\theta_r}{\theta_s-\theta_r} $
the effective saturation, where $\theta_s$ is the volumetric water content 
at saturation and $\theta_r$ is the residual volumetric water content.
We have $u\in (0,1)$, since $\theta \in (\theta_s,\theta_r)$. The soil hydraulic 
properties are represented by empirical expressions (see \cite{[vG]})
\begin{equation}
u=\frac{1}{(1+(\gamma h)^n)^m},\qquad h\in (-\infty,0)
\end{equation}
\begin{equation}
k(u)= K_s u^{1/2}[ 1 -(1-u^{1/m})^m]^2 
\end{equation}
where $m=1-1/n$, $n>1$ and $\gamma =- (2^{1/m}-1)^{1-m}/h_b$ are empirical
soil parameters, $h_b$ is called the bubbling pressure .

It is possible to rewrite the flow in unsaturated form as 
\begin{equation}
\partial_t u = \partial_r \left(D(u)\partial_r u- \frac{\omega^2}{g}k(u)r\right)
\label{eq1.3}
\end{equation}
where
\begin{multline}
D(u)=-\frac{K_s}{(n-1)\gamma (\theta_s-\theta_r)} u^{1/2-1/m}(1-u^{1/m})^{-m} \\
\times [1-(1-u^{1/m})^{m}]^2
\label{eq1.4}
\end{multline}
Equation (\ref{eq1.3}) is strongly nonlinear and degenerate. We note that
$D(0)=0, D(1)=\infty $. 
Equilibria at the high rotational speed can be expected to have a fully saturated zone (supposing the initial amount of infiltrated water is sufficiently large), which appears at the right boundary and of which the front  evolves to the left of the specimen (under  non-decreasing rotational speed). We denote the position of this interface by $s(t)$. This saturated zone is governed by Darcy's 
equation, but $s(t)$ is unknown and time dependent. The time evolution of $s(t)$ is difficult to compute. The dynamics of this region is linked with the (finite) interface flux $q_i$
$$q_i=- \left.\left(D(u)\partial_r u- \frac{\omega^2}{g}k(u)r\right)\right|_{r=s(t)},
$$
and based on a mass balance argument we can expect $$ \dot{s}(t)=-q_i.$$
Unfortunately, we cannot use this model for the determination of the time evolution of $s(t)$,  since at $r=s(t)$ it holds $u=1$ and $D(1)=\infty $. Consequently, $\partial_r u|_{r=s(t)}=0$.

\begin{figure}
\centering
    \includegraphics[width=8cm] {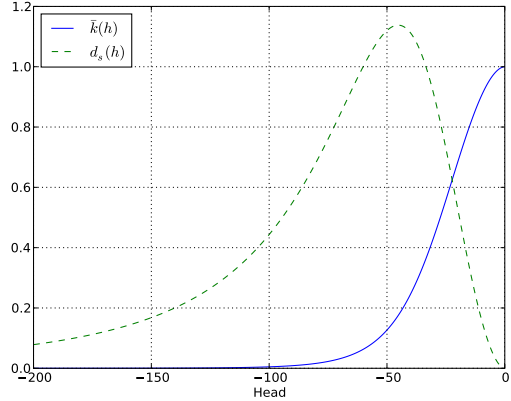}
\caption{$\bar{k}(h)$ and $100 \times d_s(h) $ for $n=2.81$,
$\gamma =-0.0189$}    
\label{fig1} 
\end{figure}

If we transform Richards' equation in terms of the head, we obtain 
\begin{equation}
d_s(h)\partial_t h =k_0 \partial_r \left[\bar{k}(h)\partial_r h- \frac{\omega^2}{g}\bar{k}(h)r\right],
\label{eq1.5}
\end{equation}
with $k_0=\frac{K_s}{\theta_s-\theta_r}$,
where $k_0 \bar{k}(h)$ is the hydraulic conductivity function,
$$\bar{k}(h)= \frac{1}{(1+(\gamma h)^n)^{m/2}}\left(1-\frac{(\gamma h)^{n-1}}{(1+(\gamma h)^n)^{m}}\right)^2, $$
and the specific moisture capacity function $d_s(h)=\mathrm{d}u/\mathrm{d}h$ is given by 
$$d_s(h)= -\gamma (n-1)\frac{(\gamma h)^{n-1}}{(1+(\gamma h)^n)^{1+m}}.$$
We can see that $\bar{k}(h)\to 1$ for $h \to 0$. In Fig.\ \ref{fig1} we present the graph of the functions $\bar{k}(h)$ and $100\, d_s(h)$ for $h\in (-200, 0)$, and parameter values $K_s=2.4\, 10^{-5}$, $n=2.81$,
$\gamma =-0.0189$.
As we can see also equation (\ref{eq1.5}) degenerates at $h=0$. This has to be taken into account when saturation becomes  $1$ at the right boundary of specimen. After this moment, $t=t_1$, the mathematical model must be changed to reflect the physical phenomenon. At the right hand side of the (isolated) specimen appears a saturated zone with an interface $s(t)$ moving from the right boundary to the left. The flux at the interface $s(t)$ is equal to $-\dot{s}(t)$, but also in this pressure-head form of Richards' equation it is  difficult to approximate correctly
$ \partial_t h|_{x=s(t)}$, which leads to a significant error in the mass balance. 

Therefore to determine the interface $s(t)$, we will consider the algebraic equation 
\begin{equation}
\int_{r_0}^{r_0+L}u\left(h(x,t)\right)\, \mathrm{d}t +L-s(t)=M_w,\quad s(0)=L,
\label{eq1.6}
\end{equation}
where $M_w$ is the amount of infiltrated water (which remains constant during the centrifugation). This condition reflects the global mass balance in the specimen and does not suffer from a flux approximation at $r=r_0+s(t)$.

Then mathematical model (\ref{eq1.5}) only needs to be solved over the interval $r\in (r_0,r_0+s(t))$ with right boundary condition $h(r_0+s(t))=0$ for all $t$. We approximate this mathematical model in the next section.

\section{Numerical method}\label{nummeth}
For the output parameters that will be measured (gravity center and rotational momentum), there is no need to model the head in the saturated zone, as we consider the compressibility of water to be negligible. The numerical approximation of mathematical model (\ref{eq1.5})-(\ref{eq1.6}) results in a coupled system of a PDE and an algebraic equation. Moreover, the solution domain is a moving region, with unknown interface $s(t)$, which has to be determined.

We shift (\ref{eq1.5}) to the domain $ r\in (0,s(t))$ and use the fixed domain transformation $ y=\frac{r}{s(t)}$. This gives
\begin{multline}
\label{eq1.7}
d_s(h)\left(\mathrm{d}_t{h}(y,t)-y \frac{\dot{s}(t)}{s(t)}\partial_y h\right) = \\
k_0\frac{1}{s(t)^2}  \partial_y \left(\bar{k}(h)\partial_y h- \bar{k}(h) \frac{\omega^2s}{g}\left(r_0+y s(t)\right) \right).
\end{multline}
Consider the space discretization
$$0=y_0< y_1<\ldots<y_i<\ldots <y_N=1,$$
and $\alpha_0=0$, $\alpha_i:=y_i-y_{i-1}$, $i=1,\ldots,N$
and integrate (\ref{eq1.7}) over $I_i:= (y_{i-1/2},y_{i+1/2})$ for $i=1,\ldots,N-1$ 
where $y_{i-1/2}:=(y_i+y_{i-1})/2$, $y_{i+1/2}:=(y_i+y_{i+1})/2$.

We denote by $h_i(t) \approx h(y_i,t)$, $\forall i=1,\ldots,N-1$, and approximate
$\mathrm{d}_t h(y,t)\approx \dot{h}_i(t)$ in the interval $I_i$. 
We approximate
$$ \partial_y h|_{y=y_{i+1/2}}\approx \frac{h_{i+1}(t)-h_i(t)}{\alpha_{i+1}}=:\partial^+h_i$$
and similarly we approximate $ \partial_yh|_{y=y_{i-1/2}}$ and denote it by 
$:\partial^-h_i$. Let ${\cal L}(z;y_i)$ be the second order Lagrange polynomial crossing
the points $(y_{i-1},h_{i-1}),(y_{i},h_{i})$ and $(y_{i+1},h_{i+1})$.
We use the abbreviation $k_{i+1/2}:= \bar{k}(h_{y_{i+1/2}})$ Then, our approximation
of (\ref{eq1.7}) (based on finite volume type approximation) at the point $y=y_i$ reads as follows
\begin{multline}\label{eq1.8}
d_s(h_i)\left(\dot{h}_i- \frac{\dot{s}y_i}{s}\left.\frac{d {\cal L}(z;y_i)}{dz}\right|_{z=y_i} \right) = \\
k_0 \frac{2}{\alpha_i+\alpha_{i+1}}\ \frac{1}{s^2} \Bigl[k_{i+1/2}  \partial^+h_i-k_{i-1/2}  \partial^-h_i-\Bigr.\\
\left.\frac{ \omega^2 s}{g}\left(k_{i+1/2}(r_0+s y_{i+1/2})-k_{i-1/2}(r_0+s y_{i-1/2})\right)\right]
\end{multline}
for $i=1,...,N-1$. We add the corresponding equation at the point $y_0$ taking into account that the flux is zero there. In a similar way as in (\ref{eq1.8}) (following the finite volume type of approximation) we obtain
\begin{multline}\label{eq1.9}
d_s(h_0)\dot{h}_0 = k_0 \frac{2}{\alpha(1)}\, \frac{1}{s^2}\\
\times
\left[k_{1/2}  \partial^+h_1 -\frac{ \omega^2 s}{g}\left(k_{1/2}(r_0+s y_{1/2})\right)\right].
\end{multline}
At the point $y_N=1$  we have $h_N(t)=0$, so no additional equation is considered.
We approximate the amount of water $M_w$ using the trapezoidal rule for the integration. Define
$$ Q(t)=\approx u_0\, \alpha_{1}/2 +\alpha_{N}/2+ \sum_1^{N-1} \frac{\alpha_{i}+\alpha_{i+1}}{2}u_i,$$
where $$u_i=\frac{1}{(1+(\gamma h_i)^n)^m}.$$
Then, system  (\ref{eq1.8})-(\ref{eq1.9}) will be completed by the algebraic equation
\begin{equation}
0 = L -s(t)[1-Q(t)] - M_w.
\label{eq1.10}
\end{equation}
This algebraic equation is used instead of an ODE equation that models $\dot{s}(t)$.
System (\ref{eq1.8})-(\ref{eq1.10}) is degenerate and is of the form
\begin{equation}
M(t,z)\dot{z}(t)= f(t,z)
\label{eq1.11}
\end{equation}
where $z=(h_0,h_1,...,h_{N-1},s)$.
The last equation of this system is just (\ref{eq1.10}). This system can be readily solved, e.g., by the solver ``ode15s'' in MATLAB$^\circledR$ or the ``ida'' solver of the Sundials package. 

As is usual with these solvers, some regularization in (\ref{eq1.10}) is needed as well as a tuning of the space discretization. Most important is to have a ``good'' starting point.

If the equilibria have the property  $h_N < 0$, then no interface appears. It is then needed to set $s(t):=L$ in the previous mathematical model and replace algebraic equation (\ref{eq1.10}) by an ODE equation for $\dot{h}_N$ which will be similar to (\ref{eq1.9}). 
Successively increasing the rotational speed of the centrifuge increases the head at the right boundary. The model remains in the state where $s(t):=L$ up to the point when $h(N) = 0$, at which point the computation is automatically halted. The full model (\ref{eq1.8})-(\ref{eq1.10}) is used onwards to compute the equilibrium states.

In the numerical equilibrium experiments (see Section \ref{numexp}, Exp.\ \ref{exp5} and \ref{exp7}) it is observed, as expected, that the values of the rotational moment $M_r$ and the center of gravity $G_c$ are not very sensitive on the important $K_s$ parameter. Also the transient experiments where the time sections between different equilibria are measured, are not very sensitive. The saturated conductivity  $K_s$ can only be determined from measurements of $M_r$, $G_c$ that are accurate up to $3$ digits. Therefore, another method must be used for the determination of $K_s$, which we present in the next section.  
 
\section{Centrifugation of saturated sample}\label{sat}
For the determination of the saturated conductivity we propose to use the device put forward in \cite{[NM]}, with the addition of allowing for transient measurements. We specifically use the ability to measure when a reservoir has completely drained out, combined with the measurements of the rotational moment.

We consider a saturated sample placed in the region $(r_0,r_0+L)$ of the centrifuge. A water reservoir is placed in front of the sample,  $r\in (r_0-l_0, r_0)$ and a head of $0$ is maintained at the bottom $r_0 + L$ via a lip (alternatively one could place an empty reservoir $r\in (r_0+L,r_0+L+d)$ into which the water flowing out from the sample is collected). The water from the front reservoir will flow through the sample and drain out to the right. We proceed the centrifugation up to the moment $T_e$ which is defined as the moment when the front reservoir is empty. By means of this setting
the measurements of rotational momentum, resp. $T_e$, will be very sensitive on the 
model parameter $K_s$. The information of only the end time $T_e$ might even be sufficient to determine $K_s$.

The mathematical model is very simple, since we have a saturated sample during the centrifugation - see (\ref{eq1.1a}), with time dependent Dirichlet boundary condition. The flux of water is given by
\begin{equation}\label{qf}
q_F(t)=-K_s\left(\partial_r h-\frac{\omega^2}{g}r\right),
\end{equation}
and is constant along all the sample. The head at the left boundary $r=r_0$ depends on the water level $\ell(t)$ (in the reservoir) and is under centrifugation given by
$$h(r_0)= \frac{\omega^2}{2\,g} \ell(t)(2r_0-\ell(t)).$$
The head at the outflow boundary is zero, i.e., $h(r_0+L)=0$. Hence, by simple integration of (\ref{eq1.1a}) in the saturated region we obtain the solution  
\begin{multline}
h(r_0+x,t)=\frac{\omega^2}{2\,g}\left[ \ell(t)(2r_0-\ell(t))+x^2\right.\\
\left.-2r_0 x\left(-1+\frac{\ell(t)+L}{L}\right)+\frac{L^2-\ell(t)^2}{2r_0L}\right],
\label{eq3.1}
\end{multline} 
where $x\in (0,L)$. Inserting (\ref{eq3.1}) in (\ref{qf}) we obtain
\begin{equation}
q_F(t)=K_s \frac{\omega^2}{2gL}\left[L^2-\ell(t)^2+ 2r_0(L+\ell(t))\right]
\label{eq3.2}
\end{equation} 
Due to the mass balance argument we obtain the governing ODE for the time evolution of the water level: 
\begin{equation}
\dot{\ell}(t)=-K_s \frac{\omega^2}{2gL}\left[L^2-\ell(t)^2+ 2r_0(L+\ell(t)))\right]\equiv -q_F(t), 
\label{eq3.3}
\end{equation} 
with $\ell(0)=l_0$ and $\ell(T_e)=0$.
Solving this ODE we obtain the relation between $T_e$ and $K_s$, whereas $\ell(t)$ fully determines the change of the rotational moment $M_r(t)$ over time. The numerical experiments for this set-up is presented in Sec.\ \ref{exp8}, below.
 
\section{Numerical experiments}\label{numexp}
For all the experiments we use as data $r_0=10$, $L=10$, $\omega=30$, $K_s= 2.4\, 10^{-5}$, $\theta_r=0.02$, $\theta_s=0.4$, $\gamma =-0.0189$, $n=2.81$, except where noted otherwise or where sequences are compared to investigate the sensitivity of the set-up on the parameters. If not mentioned otherwise, a uniformly distributed space discretization with $N=40$ grid points is used.

\subsection{Independence of initial data}\label{exp1}
\begin{figure}
\centering
    \includegraphics[width=8cm]  {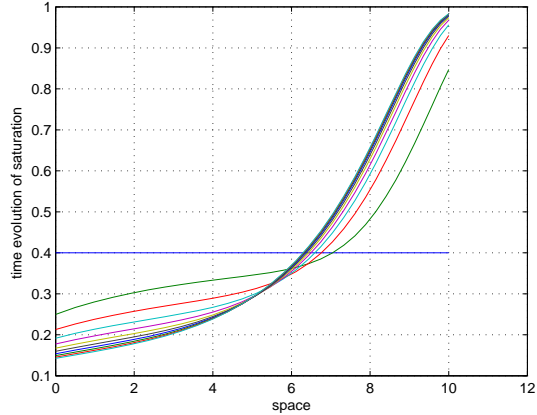}
\caption{Evolution to equilibrium starting from a uniform water distribution, $\omega =30$ }    
\label{fig2} 
\end{figure}
\begin{figure}
\centering
    \includegraphics[width=8cm]  {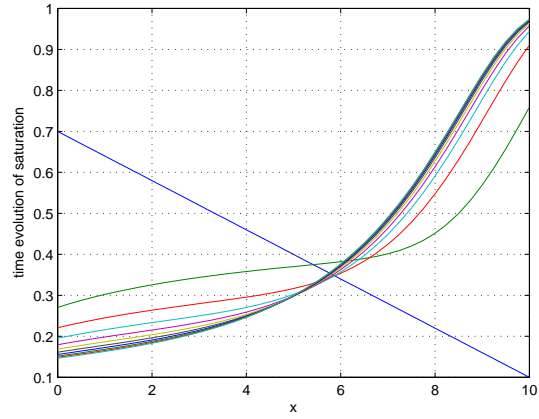}
\caption{Evolution to equilibrium starting from a lineary decreasing water distribution, $\omega =30$ }    
\label{fig3} 
\end{figure}
In the first numerical experiment we demonstrate that the equilibrium depends only on the rotational speed and the amount of infiltrated water.  
In Fig.\ \ref{fig2}  we show the time evolution of a set of unsaturated profiles over $10$ equidistant time sections in the interval $(0,500.000\mathrm{s})$. From this figure it is clear that 
at the end time the profile is very close to the equilibrium (of which the precise value 
is obtained in infinite time). Moreover, a long time before the end of the centrifugation at $t=500.000\mathrm{s}$, the saturation profiles do not change visibly anymore. 

The independence of the initial profile is clear in Fig.\ \ref{fig3} where the only change is in the starting saturation profile (which is now lineary decreasing), keeping the amount of infiltrated water identical. The same equilibrium as in Fig.\ \ref{fig2} is approached.

\subsection{Rotational speed}\label{exp2}
We investigate the sensitivity on the rotational speed $\omega$. For this we
monotonically increase $\omega$ in a stepwise way, from $\omega =10$ up to $\omega =70$ with increment $\Delta \omega = 5$, resulting in $13$ saturation profiles.  At each rotational speed we reach the corresponding ``good approximated'' equilibrium. The centrifugation from one equilibrium to the next one takes $800.000\mathrm{s}\approx 9\mathrm{days}$.
\begin{figure}
\centering
    \includegraphics[width=8cm]   {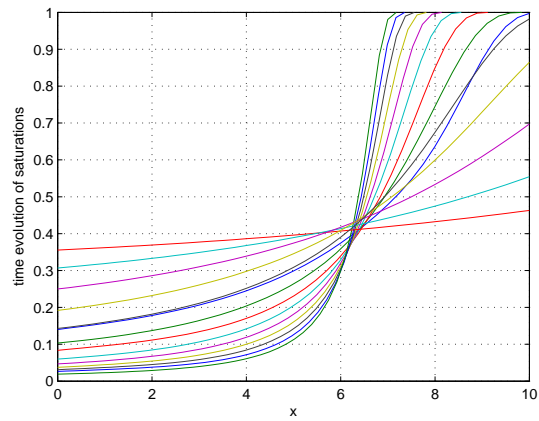}
\caption{Equilibrium saturation profiles for $\omega =10 + j\,5$, $j=0,\ldots,12$, and profile with  $h_N=0$}    
\label{fig4} 
\end{figure}
\begin{figure}
\centering
    \includegraphics[width=8cm]  {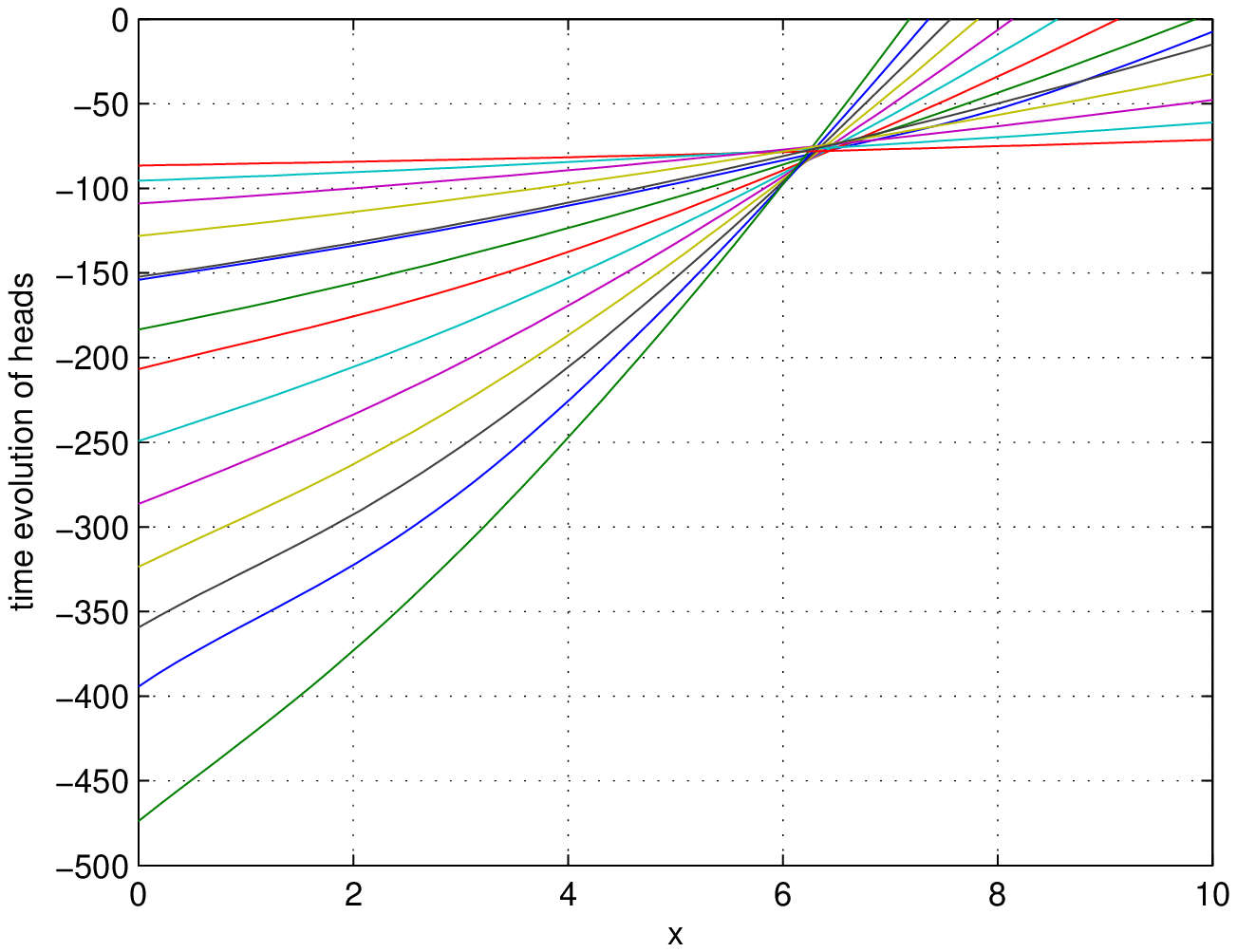}
\caption{Equilibrium head profiles for $\omega =10 + j\,5$, $j=0,\ldots,12$, and profile with  $h_N=0$}    
\label{fig5} 
\end{figure}
In Fig. \ref{fig4} we present the saturation profiles of the equilibria, together with one extra profile corresponding to the right boundary becoming fully saturated. The corresponding heads are presented in Fig.\ \ref{fig5}. 

The extra profile is the sixth curve, which occurs during $\omega=35$ and  has value $h_N=0$. At that moment, the ODE solver is stopped, the system 
is changed to the system with a moving boundary, Eq.\ (\ref{eq1.5})-(\ref{eq1.6}), and the solver is restarted with initial condition given by the last potential head (resp. saturation) profile.

The values at the equilibria of the rotational momentum $M_r$, the  center of gravity $G_c$, and total water content $M_w$  are collected in Table \ref{tab1}. These are computed from the saturation profiles.
\begin{table}
\caption{Rotational momentum, center of gravity, water amount for Exp.\ \ref{exp2}}
\label{tab1}
\begin{tabular}{|l|l|l|l|l|}
\hline $\omega$ & $M_{r,s}\,10^{-6}$&$M_{r,e}\,10^{-6}$ & $G_c$ & $M_w$ \\
\hline 10 & 0.0467 & 0.0481 &4.9999&3.9999\\
\hline 15 & 0.1081& 0.1126& 5.4979&4.0020\\
\hline 20 & 0.2002 &0.2118 & 5.8722&4.0041\\
\hline 25 & 0.3309 & 0.3498 & 6.3012&4.0057\\
\hline 30 & 0.5037 & 0.5288& 6.7097&4.0087\\
\hline 35 & 0.7037 & 0.72307& 6.8220&4.0110\\
\hline 40 & 0.9307 & 0.9747& 7.2727&4.0141\\
\hline 45 & 1.2338 & 1.2534& 7.4323&4.0137\\
\hline 50& 1.5474 & 1.5649& 7.5494&4.0133\\
\hline 55 & 1.8936 & 1.9089& 7.6349&4.0130\\
\hline 60 & 2.2718 & 2.2853& 7.6984&4.0126\\
\hline 65 &2.6821  &2.6941 &7.7469 &4.0123\\
\hline 70 & 3.1245 & 3.1352&7.77845 &4.0120\\
\hline
\end{tabular}
\end{table}
Two rotational moments are given, $M_{r,s}$ denotes the starting rotational momentum corresponding to the saturation distribution obtained from the previous rotational speed, and $M_{r,e}$ denotes the rotational momentum at the actual rotational speed in equilibrium. The formulas for $M_r$, $G_c$ and $M_w$ at time $t$ are:
$$ M_r = \frac{ s(t)}{2}\int_0^1 (r_0+s(t)z)^2u(t,z)dz+ \frac{1}{6}(L^3-s(t)^3),$$ $$M_w=s(t)\int_0^1u(t,z)dz+\frac{1}{2}(L^2-s(t)^2),$$
$$G_c=s(t)\int_0^1yu(t,z)dz/M_w,$$
and are all evaluated numerically using the trapezoidal rule. Note that if $u(t,1)<1$ then $s(t)=L$.
The sensitivity of the measured quantities on the changing water content is very good. It remains to determine the contribution of the different parameters.

We note first however that careful analysis of the head profiles in Fig.\ \ref{fig5} shows that these are not yet parabola at the end times of a run at fixed rotational speed, especially toward the left of the sample. This indicates that the equilibrium has not yet been reached. We investigate this further in the next Experiment.

\subsection{Reaching equilibrium}\label{exp3}

\begin{figure}
\centering
    \includegraphics[width=8cm]  {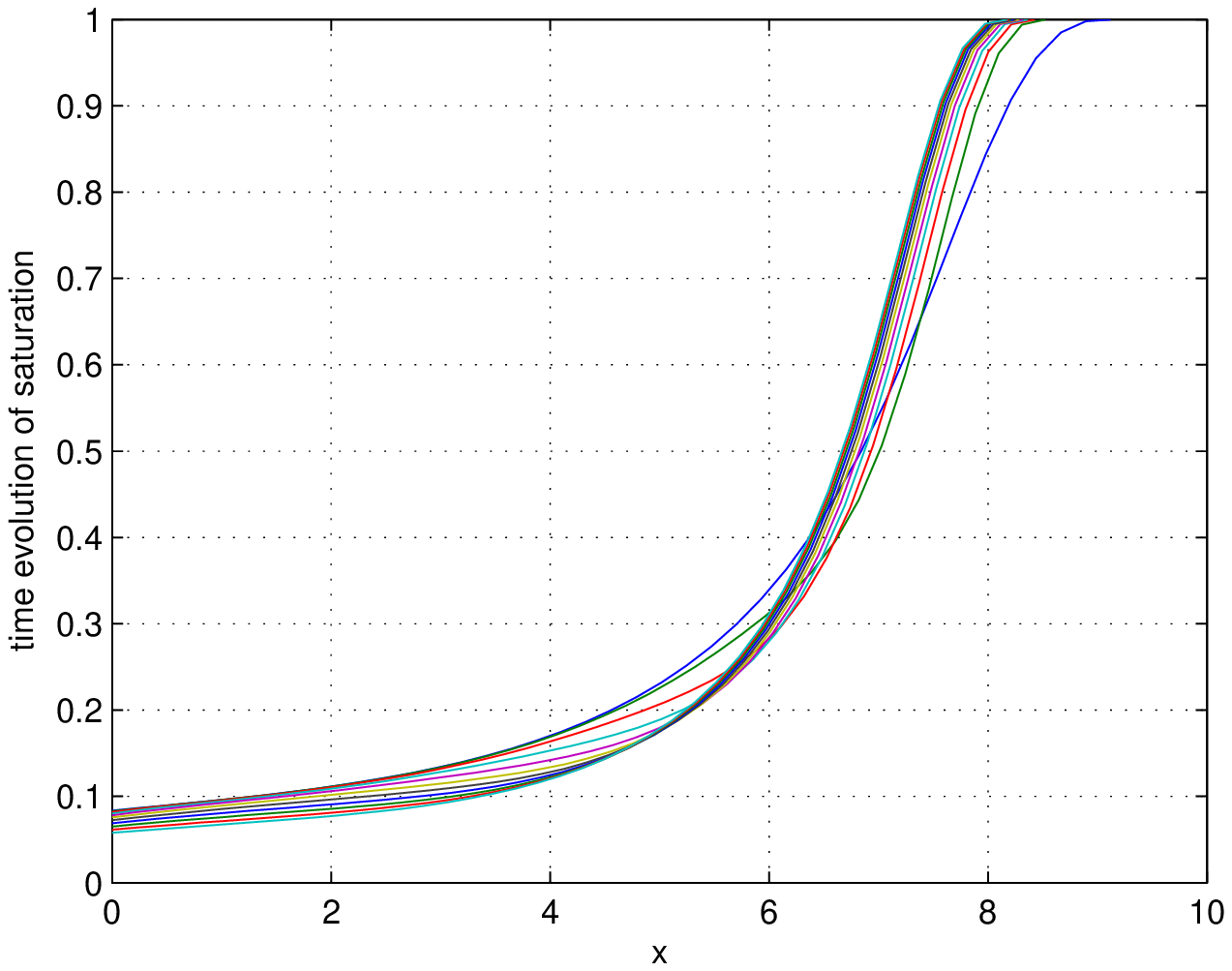}
\caption{Evolution of the saturation starting from equilibrium at $\omega =40$, with $\omega =50$ }    
\label{fig6} 
\end{figure}

\begin{figure}
\centering
    \includegraphics[width=8cm]  {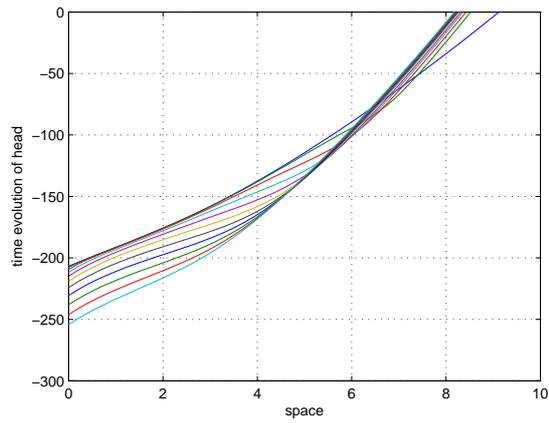}
\caption{Evolution of potential head corresponding with Fig.\ \ref{fig6}}
\label{fig7} 
\end{figure}

To investigate the head profiles we start in this experiment from the equilibrium corresponding to $\omega=40$ and use a rotational speed $\omega=50$. The centrifuge normally operates up to $T_e=1.540.000$ seconds. At that time, equilibrium for $\omega=50$ is almost reached. We compare $13$ values, the starting value, 9 increasing time steps (with $\Delta t_j=t_{j+1}-t_j= 2000\, 2^j,\ j=1,...,9$), the sensible end time step $T_e=770 \times 2.000\mathrm{s}$, and 2 extra time steps to investigate the very long time behavior. Fig.\ \ref{fig6} shows the obtained saturation profiles, the corresponding evolution of the potential head is presented in Fig.\ \ref{fig7}.
\begin{table}
\caption{Rotational momentum, center of gravity, water amount for Exp.\ \ref{exp3}}
\label{tab2}
\begin{tabular}{|l|l|l|l|}
\hline $\frac{\mathrm{time}}{2000}\mathrm{s}$ &$M_{r,e}\,10^{-6}$ & $G_c$ & $M_w$ \\
\hline 0& 1.5201 & 7.2512  & 4.0141\\
\hline 1 & 1.5248  &  7.2813 &   4.0128\\
\hline 3 &1.5299  &  7.3119  &  4.0131 \\
\hline 7 & 1.5345  &  7.3413  &  4.0133\\
\hline 15& 1.5389  &  7.3697  &  4.0134\\
\hline 31& 1.5430   & 7.3972   & 4.0134\\
\hline 63 & 1.5469  &  7.4234  &  4.0135\\
\hline 127 & 1.5505  &  7.4478   & 4.0135\\
\hline 255& 1.5537   & 7.4699  &  4.0136\\
\hline 511& 1.5565   & 7.4893  &  4.0136\\
\hline 770 &  1.5588  &  7.5056 &   4.0136\\
\hline 1800 & 1.5645  & 7.5462 & 4.0133\\
\hline 2300 &  1.5649 & 7.5494 & 4.0133\\
\hline
\end{tabular}
\end{table}
As we can see the head profile at $T_e$ is still not a parabola in the left side of the profile (low head values). The reason for this is that the hydraulic permeability at low head is negligibly small, so it takes a very long time to reach the equilibrium. If the centrifugation is continued, also this part obtains the required parabolic shape associated with the equilibrium. Note however, that the other part of the head profile (for higher head values) is changing insignificantly. Therefore, we arrive at the conclusion that it makes sense to increase the rotational speed and not wait for these lower head values to stabilize. More so in light of Exp.\ \ref{exp1}.

The measured values for the rotational momentum, gravity center, and water amount, are given in Table \ref{tab2}. The small change between the last two values in Table \ref{tab2} demonstrates that equilibrium is eventually reached.

\subsection{Dependence on $n$}\label{exp4}
\begin{figure}
\centering
    \includegraphics[width=8cm]  {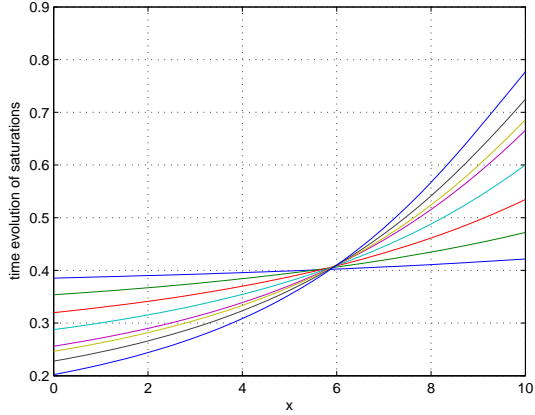}
\caption{Equilibrium saturation profiles at $\omega =20$ for 
  $n=1.51$; $1.81$; $2.11$; $2.41$;  $2.71$; $2.81$; $3.01$ and $3.31$.}
\label{fig8} 
\end{figure}

\begin{figure}
\centering
    \includegraphics[width=8cm]  {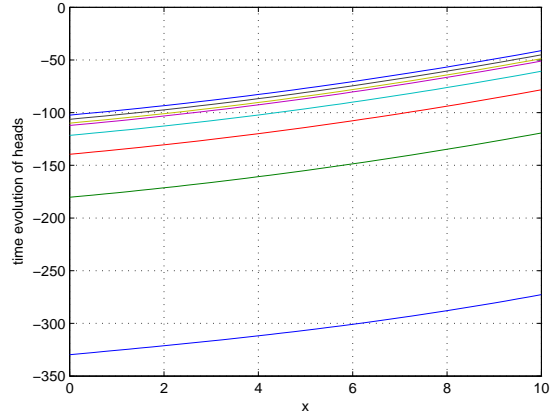}
\caption{Equilibrium head profiles corresponding with Fig.\ \ref{fig8}}
\label{fig9} 
\end{figure}

In this experiment we demonstrate the sensitivity of $M_r$ and $G_c$ on the model parameter $n$. We start with a constant saturation $u=0.4$ and apply now the rotational speed $\omega=20$. The centrifuge is operated for $800.000\mathrm{s}$. In Fig.\ \ref{fig8} the obtained equilibrium profiles are depicted for successively $n=1.51$; $1.81$; $2.11$; $2.41$;  $2.71$; $2.81$; $3.01$ and $3.31$. In Fig.\ \ref{fig9} the corresponding heads are given.

The resulting values for  $M_r$, $G_c$ and $M_w$ are given in Table \ref{tab3}, and indicate a good sensitivity.
\begin{table}
\caption{Rotational momentum, center of gravity and water amount for Exp.\ \ref{exp4}}
\label{tab3}
\begin{tabular}{|l|l|l|l|}
\hline $n$ &$M_re .10^{-6}$ & $G_c$ & $M_w$ \\
\hline 1.51& 0.1887 & 5.0736  & 4.0043\\
\hline 1.81 & 0.1927 &  5.2391 &   4.0052\\
\hline 2.11 &0.2020  &  5.6188  &  4.0043 \\
\hline 2.41 &0.2068   &  5.8096  &  4.0062\\
\hline 2.71& 0.2083  &  5.8701 &  4.0054\\
\hline 2.81& 0.2112  & 5.9870   & 4.052\\
\hline 3.01 & 0.2153  &  6.1524 &  4.0060\\
\hline 3.31 & 1.5505  &  7.4478   & 4.0135\\
\hline
\end{tabular}
\end{table}

\subsection{Saturated conductivity}\label{exp5}
We now proceed with investigating the sensitivity of $M_r$ and $G_c$ on the saturated conductivity $K_s$.
We consider $K_s=k_s\, 10^{-5}$ for $k_s=0.4$; $0.8$; $1.6$; $2.0$; $2.4$; $2.8$ and $3.2$,  start again from a uniform saturation, and apply this time a rotational speed of  $\omega=20$. In Table \ref{tab4} we present the resulting values of $M_r$, $G_c$ and $M_w$
at the obtained equilibrium.
\begin{table}
\caption{Rotational momentum, center of gravity and water amount for Exp.\ \ref{exp5}, $\omega=20$}
\label{tab4}
\begin{tabular}{|l|l|l|l|}
\hline $k_s$ &$M_{r,e} .10^{-6}$ & $G_c$ & $M_w$ \\
\hline 0.4& 0.2049 & 5.7286  & 4.0052\\
\hline 0.8 & 0.2077 &  5.8401 &   4.0078\\
\hline 1.2 &0.2081  &  5.8633  &  4.0059 \\
\hline 1.6 &0.2087   &  5.8684  &  4.0098\\
\hline 2.0& 0.2082  &  5.8697 &  4.0074\\
\hline 2.4& 0.2083  & 5.8700  & 4.054\\
\hline 2.8 & 0.2083  &  5.8703 &  4.0062\\
\hline 3.2 & 0.2087 &  5.8702   & 4.0095\\
\hline
\end{tabular}
\end{table}
The sensitivity of $M_r$ and $G_c$ on $k_s$ is negligible at the relative low pressure head that is generated for $\omega=20$. 

In Table \ref{tab5} we present analogous results for the equilibrium obtained at rotational speed $\omega=50$, starting from the equilibrium
at $\omega=40$, and considering a much larger spectrum for $k_s$.
\begin{table}
\caption{Rotational momentum, center of gravity and water amount for Exp.\ \ref{exp5}, $\omega=50$}
\label{tab5}
\begin{tabular}{|l|l|l|l|}
\hline $k_s$ &$M_{r,e}\, 10^{-6}$ & $G_c$ & $M_w$ \\
\hline 0.0001& 0.1867 & 5.0118  & 4.0006\\
\hline 0.001 & 0.1888 &  5.0855 &   4.0025\\
\hline 0.01 &0.1998  &  5.5221  &  4.0059 \\
\hline 0.1 &0.2085   &  5.8697  &  4.0063\\
\hline 1.0& 0.2084  &  5.8699 &  4.0074\\
\hline 10& 0.2083  & 5.8701  & 4.0064\\
\hline 100 & 0.2083  &  5.8703 &  4.0022\\
\hline
\end{tabular}
\end{table}
To reach the equilibrium from which Table \ref{tab5} is extracted, a centrifugation time of $15\,10^6\mathrm{s}$ is used for the values $k_s<1$, whereas for values  $k_s>1$ we have used an operating time of $10^6\mathrm{s}$.

\begin{figure}
\centering
    \includegraphics[width=8cm]   {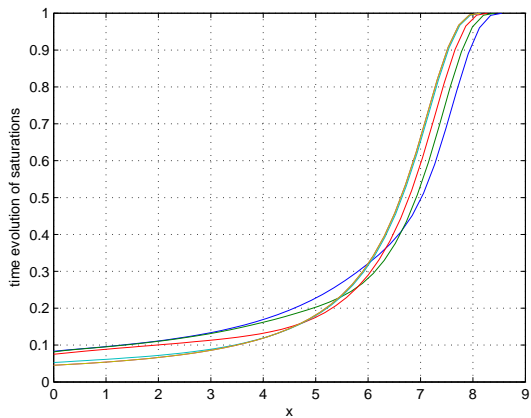}
\caption{evolution of saturations at equilibrium for $\omega =50$,
 $k_s=0.0001;\ 0.001;\ 0.01;\ 0.1;\ 1;\ 10;\ 100 $}    
\label{fig10} 
\end{figure}

\begin{figure}
\centering
    \includegraphics[width=8cm]  {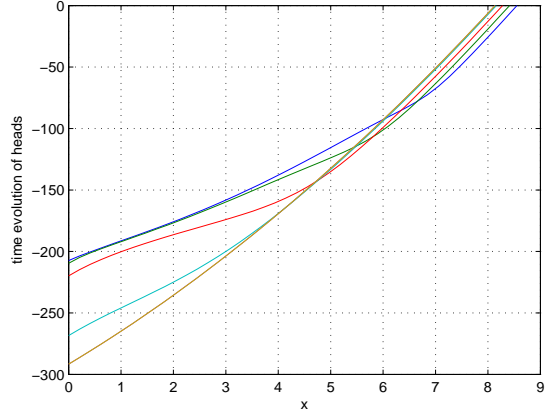}
\caption{evolution of heads at equilibrium for $\omega =50$,
 $k_s=0.0001;\ 0.001;\ 0.01;\ 0.1;\ 1;\ 10;\ 100 $}    
\label{fig11} 
\end{figure}

In Fig.\ \ref{fig10} the equilibrium profiles of the saturation is given corresponding with the results from Table \ref{tab5}, while the corresponding heads are given in Fig.\ \ref{fig11}. From Fig.\ \ref{fig11} one can see that for the lower values of the head a parabolic shape of the profile was still not reached if $k_s<1$.
Overall, the results clearly indicate that another technique must be used to determine $K_s$.

\subsection{Dependence on $\gamma$}\label{exp6}
\begin{figure}
\centering
    \includegraphics[width=8cm]  {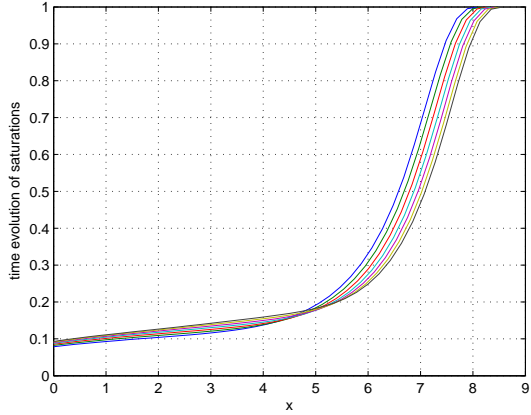}
\caption{Saturation profiles at equilibrium for $\omega =50$,
 $\gamma_0=1.59;\ 1.69;\ 1.79;\ 1.89;\ 1.99;\ 2.09;\ 2.19$}    
\label{fig12} 
\end{figure}

\begin{figure}
\centering
    \includegraphics[width=8cm]  {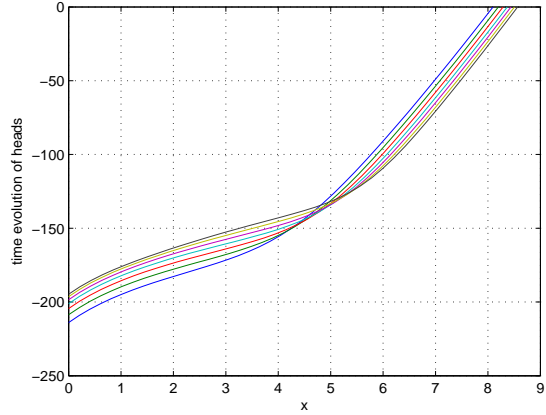}
\caption{Head profiles corresponding with Fig.\ \ref{fig12}.}    
\label{fig13} 
\end{figure}

We now investigate the sensitivity of $M_r$ and $G_c$ on the last model parameter: $\gamma$. We again use a rotational speed of  $\omega=50$, starting from the equilibrium position at $\omega=35$. As values for $\gamma$ we consider $\gamma=-\gamma_0\,10^2$  with $\gamma_0 \in (1.59;\ 2.19)$ where increments of size $0.1$ are used. The values of $M_r$ and $G_c$  are listed in Table \ref{tab6} and  Table \ref{tab6a}, respectively. The corresponding saturation and head profiles at time section $t=10^5$ are in Fig.\ \ref{fig12} and Fig.\ \ref{fig13}, respectively. 
\begin{table*}
\caption{Rotational momentum $M_r\,10^{-6}$ for Exp.\ \ref{exp6},}
\label{tab6}
\begin{tabular}{|l|l|l|l|l|l|l|l|}
\hline time{$\backslash$}$\gamma_0$ & 1.59& 1.69& 1.79& 1.89& 1.99& 2.09 & 2.19 \\
\hline \hline 1000& 1.5189 &1.5093&1.5013 &1.4949 &1.4896& 1.4854& 1.4819\\
\hline 3000 &1.5352&1.5213&1.5097 & 1.5000& 1.4919& 1.4852&  1.4797\\
\hline 5000 &1.5438&1.5278&1.5144  &1.5030 & 1.4935&1.4855 & 1.4788\\
\hline 10$^4$ &1.5565 &1.5376&1.5216 &1.5079 & 1.4963&1.4864 & 1.4780\\
\hline 5.10$^4$&1.5901 &1.5645&1.5423& 1.5231& 1.5063&1.4917 & 1.4791\\
\hline 10$^5$& 1.6058 &1.5777&1.5530  &1.5313& 1.5124&1.4958 &  1.4812\\
\hline
\end{tabular}
\end{table*}
\begin{table*}
\caption{Center of gravity for Exp.\ \ref{exp6}}
\label{tab6a}
\begin{tabular}{|l|l|l|l|l|l|l|l|}
\hline time{$\backslash$}$\gamma_0$ & 1.59& 1.69& 1.79& 1.89& 1.99& 2.09 & 2.19 \\
\hline \hline 1000&7.1117& 7.1031& 7.0948& 7.0872& 7.0805& 7.0744& 7.0693 \\
\hline 3000&7.1487& 7.1379& 7.1271& 7.1170& 7.1076& 7.0990& 7.0912\\
\hline 5000&7.1697& 7.1581& 7.1464& 7.1351& 7.1245& 7.1146& 7.1056\\
\hline 10$^4$&7.2026& 7.1903& 7.1775& 7.1649& 7.1527& 7.1411& 7.1303\\
\hline 5.10$^4$&7.3012& 7.2898& 7.2763& 7.2617& 7.2466& 7.2314& 7.2167\\
\hline 10$^5$&7.3512 & 7.3429 &7.3309 &7.3167 &7.3012 &7.2851 &7.2689 \\
\hline
\end{tabular}
\end{table*}
In Tables \ref{tab6}  and \ref{tab6a} the water amount is $4.05$.
The sensitivity on $\gamma$  is less than that of $n$, but is sufficient. Nevertheless, taking transient information into account, as given in the rows of Tables \ref{tab6}  and \ref{tab6a}, will benefit the determination of $\gamma$ via this experimental set-up.

\subsection{Transient saturation determination}\label{exp7}
In experiment \ref{exp5} we have shown the low sensitivity of $M_r$ on $K_s$ in an equilibrium analysis. One can assume this might improve in a transient analysis, as flow around the interface will be determined by $K_s$. We investigate this now.

To assure the sample is in part fully saturated, high rotational speeds are used. At fixed $K_s$ we start from the equilibrium saturation at $\omega =35$ and let 
the centrifuge operate for $10^5 \mathrm{s}$ ($\approx 27.8$hours) with $\omega = 40$. After this time we change the rotational speed with the increment $\Delta \omega=5$ and again operate for the same duration. This is continued up to rotational speed $\omega =60$, leading to $5$ end starting times $j\, 10^5 $, $j=0\ldots, 4$ and five end times $j\, 10^5$, $j=1,\ldots,5$. 

The same procedure is applied for a series of $K_s$ values. Then we compare the values of $M_r$ and $G_c$ at the same time levels for different $K_s$. In Table \ref{tab7} we present the values of $M_r$ at time steps $t_j=1000+j\,10^5$, $j=0,\ldots,4$ (index $j$ corresponds to the rows). The $9$ columns correspond to the values $K_s = k_s\, 10^{-5}$ with $k_s= 0.24$; $0.78$; $1.32$; $1.86$; $2.4$; $7.8$; $13.2$; $18.6$ and $24$.
\begin{table*}
\caption{Rotational momentum vs. $k_s$ under changing $\omega$ at the time level $t_j=1000+j\,10^5$}
\label{tab7}
\begin{tabular}{|l|l|l|l|l|l|l|l|l|l|}
\hline $t_j$ &$k_1$ & $k_2$ & $k_3$ &$k_4$ &$k_5$ &$k_6$ &$k_7$ &$k_8$ &$k_9$  \\
\hline $t_0$& 0.9526& 0.9531& 0.9536& 0.9538& 0.9540& 0.9555& 0.9564& 0.9570& 0.9576 \\
\hline $t_1$ & 1.2123 &1.2169 &1.2196 &1.2213 &1.2228 &1.2305 &1.2339 &1.2358 &1.2370\\
\hline $t_2 $&1.5049 &1.5147 &1.5200 &1.5237 &1.5265 &1.5401 &1.5450 &1.5475 &1.5490\\
\hline $t_3$ &1.8301 &1.8458 &1.8540 &1.8597 &1.8640 &1.8824 &1.8884 &1.8914 &1.8933 \\
\hline $t_4$& 2.1876 &2.2100 &2.2213 &2.2289 &2.2346 &2.2571 &2.2642 &2.2677 &2.2698\\
\hline
\end{tabular}
\end{table*}
So, in this table $M_r$  is computed $1000$s  after each change of $\omega$.
The values of $M_r$ at time level $t_{j}=j\, 10^5$, $j=1,\ldots,5$ after each change of $\omega$ is presented in Table \ref{tab8}.
\begin{table*}
\caption{Rotational momentum vs. $k_s$ under changing $\omega$ at the time level $t_{j}=j\,10^5$}
\label{tab8}
\begin{tabular}{|l|l|l|l|l|l|l|l|l|l|}
\hline $t_{j}$ &$k_1$ & $k_2$ & $k_3$ &$k_4$ &$k_5$ &$k_6$ &$k_7$ &$k_8$ &$k_9$  \\
\hline $t_1$& 0.9578 & 0.9610 &   0.9629 & 0.9641 & 0.9651 & 0.9702 & 0.9723  &0.9734 & 0.9740 \\
\hline $t_2$ &1.2191& 1.2266 & 1.2307  & 1.2334  & 1.2356  & 1.2457 & 1.2492 & 1.2509  & 1.2518 \\
\hline $t_3$&1.5127 & 1.5253 & 1.5319 & 1.5364 & 1.5398  & 1.5542 & 1.5587  & 1.5609   & 1.5622\\
\hline $t_4$ &1.8386 &1.8570 & 1.8663 &  1.8725 & 1.8771 & 1.8953 & 1.9009 &  1.9036 &  1.9052 \\
\hline $t_5$& 2.1968 & 2.2215 & 2.2337 &  2.2416  & 2.2473 &  2.2690 & 2.2755  & 2.2786 &2.2806\\
\hline
\end{tabular}
\end{table*}

In the Tables \ref{tab9} and \ref{tab10} we present the values of $G_c$ which correspond to the values of $M_r$ in Tables \ref{tab7} and \ref{tab8} (at the identical location).
\begin{table*}
\caption{Center of gravity vs. $k_s$ under changing $\omega$ at the time level $t_j=1000+j\,10^5$}
\label{tab9}
\begin{tabular}{|l|l|l|l|l|l|l|l|l|l|}
\hline $t_j$ &$k_1$ & $k_2$ & $k_3$ &$k_4$ &$k_5$ &$k_6$ &$k_7$ &$k_8$ &$k_9$  \\
\hline $t_0$&7.0485 & 7.0532 & 7.0580 & 7.0591 & 7.0608 & 7.0745 & 7.0826 & 7.0885 & 7.0946 \\
\hline $t_1$ &7.0988 & 7.1342 & 7.1554 & 7.1694 & 7.1810 & 7.2444 & 7.2722 & 7.2877 & 7.2975\\
\hline $t_2$ &7.1515 & 7.2133 & 7.2474 & 7.2721 & 7.2912 & 7.3821 & 7.4154 & 7.4324 & 7.4427\\
\hline $t_3$ &7.2011 & 7.2840 & 7.3288 & 7.3600 & 7.3836 & 7.4866 & 7.5206 & 7.5376 & 7.5480\\
\hline $t_4$& 7.2462 & 7.3460 & 7.3981 & 7.4335 & 7.4596 & 7.5660 & 7.5993 & 7.6159 & 7.6261\\
\hline
\end{tabular}
\end{table*}
\begin{table*}
\caption{Center of gravity vs. $k_s$ under changing $\omega$ at the time level $t_{j}=j\,10^5$}
\label{tab10}
\begin{tabular}{|l|l|l|l|l|l|l|l|l|l|}
\hline $t_{j}$ &$k_1$ & $k_2$ & $k_3$ &$k_4$ &$k_5$ &$k_6$ &$k_7$ &$k_8$ &$k_9$  \\
\hline $t_1$&7.0965 & 7.1279 & 7.1469 & 7.1591 & 7.1693 & 7.2235 & 7.2462 & 7.2579 & 7.2644\\
\hline $t_2$&7.1496 & 7.2085 & 7.2419 & 7.2644 & 7.2823 & 7.3672 & 7.3971 & 7.4115 & 7.4197\\
\hline $t_3$ &7.1993 & 7.2802 & 7.3238 & 7.3540 & 7.3769 & 7.4757 & 7.5073 & 7.5225 & 7.5315\\
\hline $t_4$ &7.2447 & 7.3429 & 7.3941 & 7.4288 & 7.4544 & 7.5578 & 7.5894 & 7.6048 & 7.6140\\
\hline $t_5$&7.2859  & 7.3977 & 7.4539 & 7.4910 & 7.5178 & 7.6210 & 7.6518 & 7.6671 & 7.6763\\
\hline
\end{tabular}
\end{table*}

Although the changes in the tables are clear and consistent, they remain low. Tables \ref{tab8} and \ref{tab10} effectively correspond to the equilibrium measurements, whereas the comparison with Tables \ref{tab7} and \ref{tab9} gives an estimate of what is gained by a transient analysis. We can only conclude that for this set-up also a transient analysis to determine $K_s$ adds little improvements to the sensitivity. Note that for the parameters where an equilibrium analysis results in good sensitivity, a transient analysis is possible and can dramatically decrease the centrifuge operating time.

\subsection{Saturated conductivity - 2}\label{exp8}
\begin{figure}
\centering
    \includegraphics[width=8cm]   {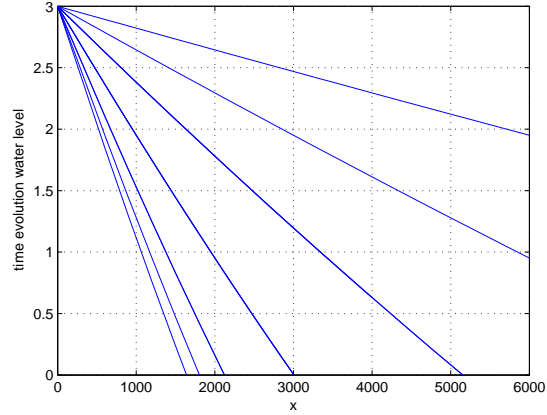}
\caption{Evolution of the water level in the reservoir, $k_s= 0.5$; $1$; $1.4$; $2.4$; $3.4$ and $4.4$}    
\label{fig14} 
\end{figure}

\begin{figure}
\centering
    \includegraphics[width=8cm]  {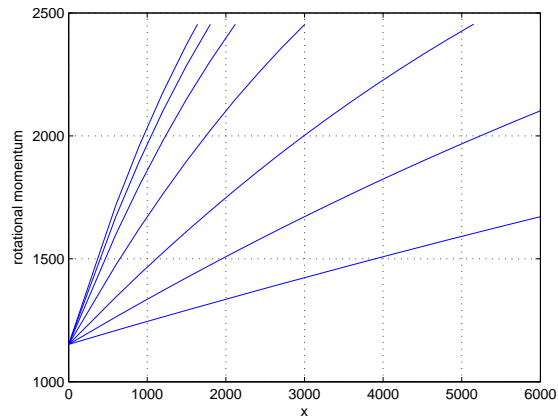}
\caption{Evolution of the rotational momentum, $k_s= 0.5$; $1$; $1.4$; $2.4$; $3.4$ and $4.4$} 
\label{fig15} 
\end{figure}

In this experiment we demonstrate that the determination of model parameter $K_s$ can be easily realized by means of very simple measurements obtained by centrifugation of a fully saturated sample with a water reservoir as presented in Section \ref{sat}. In Fig.\ \ref{fig14} and \ref{fig15} the time evolution of the water level $\ell(t)$ and of the rotational momentum $M_r(t)$ is given. Note that in the computation of $M_r(t)$ also a water collector placed at $r\in (r_0+L,r_0+L+d)$ is considered. The parameter values considered are $K_s=k_s\, 10^{-5}$ with $k_s= 0.5$; $1$; $1.4$; $2.4$; $3.4$ and $4.4$. The sensitivity of $M_r$ on $k_s$ will increase with higher values of $\omega$. 

In \cite{[NM]} $\ell(t)$ for a fixed  $t\in [0, T_e]$ is used for the determination of $K_s$. From Fig.\ \ref{fig14} and \ref{fig15} we can deduce that also the values of $M_r(t)$ (at a fixed $t\in [0, T_e]$ ) can be used for the determination of $K_s$. Adding transient data makes this determination still quicker.

\section{Acknowledgements*}
The first and the third author confirm financial support by the Slovak Research and Development Agency under contract APVV-0351-07.

\bibliographystyle{plain}

\bibliography{article_04}   

\end{document}